\documentclass[showpacs,amsmath,amssymb,superscriptaddress,aps,a4paper,prl,twocolumn]{revtex4-1}
\usepackage{graphicx}
\usepackage{dcolumn}
\usepackage{bm}
\usepackage{textcomp}
\usepackage{color}

\begin{document}

\preprint{APS/123-QED}

\title{High-contrast Aharonov-Bohm oscillations in the acoustoelectric transport regime}

\author{E.~Strambini}
\email{e.strambini@utwente.nl}
\altaffiliation{Now at University of Twente, Twente, The Netherlands}
\author{E.~Ligar\`o}
\author{F.~Beltram}
\affiliation{NEST, Scuola Normale Superiore and Istituto Nanoscienze CNR, Piazza San Silvestro 12, 56127 Pisa, Italy}

\author{H.~Beere}
\author{C.~A.~Nicoll}
\author{D.~A.~Ritchie}
\affiliation{Cavendish Laboratory, University of Cambridge, Cambridge CB3 0HE, United Kingdom}

\author{V.~Piazza}
\email{v.piazza@sns.it}
\altaffiliation{Now at Center for Nanotechnology Innovation @NEST, Istituto Italiano di Tecnologia, Piazza San Silvestro 12, I-56127 Pisa, Italy} 
\affiliation{NEST,  Scuola Normale Superiore and Istituto Nanoscienze CNR, Piazza San Silvestro 12, 56127 Pisa, Italy}

\date{\today}

\begin{abstract}
Phase-coherent acoustoelectric transport is reported.Aharonov-Bohm oscillations in the acoustoelectric current with visibility exceeding 100\% were observed in mesoscopic GaAs rings as a function of an external magnetic field at cryogenic temperatures. A theoretical analysis of the acoustoelectric transport in ballistic devices is proposed to model experimental observations. Our findings highlight a close analogy between acoustoelectric transport and thermoelectric properties in ballistic devices.
\end{abstract}


\maketitle

The pioneering work by Aharonov and Bohm had far-reaching fundamental implications even beyond its objective of highlighting the significance of potential with respect to force in quantum mechanics. 
Aharonov-Bohm (AB) interference \cite{aharonov_bohm_59} was observed in several mesoscopic systems \cite{washburn_86, ismail_91, pedersen_00, piazza_00} and exploited or proposed as a tool to investigate electron coherent dynamics at the nanoscale \cite{schuster_97, vanderwiel_00, strambini_chirolli_10}. Unfortunately its implementation in \textit{practical} devices is hindered by the rather low contrast of the conductance oscillations achievable in solid-state devices. Even at sub-K temperatures in the linear transport regime, contrast is typically limited to few percents of the background signal owing to the sizable fraction of electrons that propagate incoherently. 

In this Letter we investigate electronic phase coherence in mesoscopic AB rings in the acoustoelectric-transport regime, i.e. when electrons propagate owing to piezoelectric interactions between acoustic waves traveling on the device surface and conduction-band electrons. Our data and analysis highlight the interplay between coherence and acoustoelectric transport.

\textit{Very-high-contrast} AB oscillations will be shown in the acoustoelectric current, even if the characteristic coherence length in this transport regime will be shown to be comparable to that of the linear-regime case.  
The theoretical analysis presented here to describe the experimental investigation demonstrates that, at cryogenic temperatures and in the energy range here of interest, the acoustoelectric current shows the same behavior of a current generated by a thermal gradient across the device. Our results establish a close link between acoustoelectric and thermoelectric transport properties and open the way to the study of thermoelectric effects in nanodevices with the much simpler approach characteristic of acoustoelectric experiments.

\begin{figure}[t]
\begin{center}
\includegraphics[width=8cm]{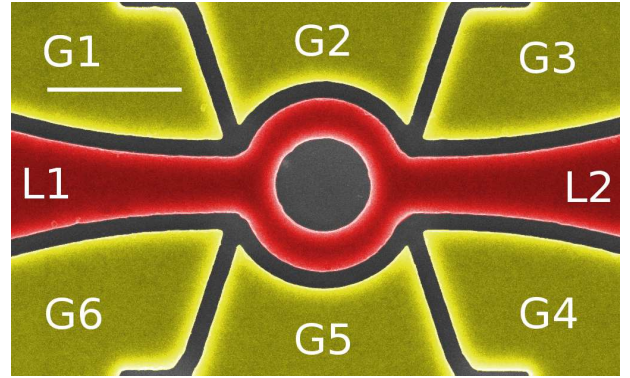}
  \caption{(Color online) SEM image of the AB ring of 500 nm radius (R500).
$L_1$ and $L_2$ label the two leads of the device. Control of the electron density in the different parts of the devices was achieved by means of the six gates $G_j$ ($j=1,...,6$).} \label{fig1}
\end{center}
\end{figure}

AB rings were fabricated starting from a two-dimensional electron gas (2DEG) confined 90 nm below the surface of a modulation-doped GaAs/Al$_{0.3}$Ga$_{0.7}$As heterostructure. At a temperature $T = 4.2$ K the unpatterned 2DEG density and mobility were $2.1 \times 10^{11}$ cm$^{-2}$ and $1.7 \times 10^6$ cm$^2$/Vs, respectively.
The ring geometry was defined by shallow plasma etching. The same processing step yielded a set of lateral gates
(labeled G$_1$ through G$_6$) that provide control over the electron density in each part of the device. 
A scanning electron microscopy (SEM) image in artificial colors of one of our devices is shown in Fig.~\ref{fig1}. 
Standard Ni/AuGe/Ni/Au (5 nm/180 nm/5 nm/100 nm) $n$-type Ohmic contacts (not shown in the figure) were fabricated to allow electrical access to the 2DEG. 
Aluminum interdigitated transducers (IDT) were evaporated to generate surface acoustic waves (SAWs). Devices with ring radii of 500 nm (device R500) and 750 nm (device R750) and IDTs tuned at resonance frequencies of 1.5 and 3 GHz (corresponding to SAW wavelengths of 2 and 1 $\mu$m) were fabricated and studied in a $^3$He cryostat at a base temperature of 350 mK.

The coherent-transport properties of the devices were first assessed in the linear regime. The injection contact (lead L$_1$) was biased with an ac excitation signal ($V_{ex}$) at 15.7 Hz with an amplitude of $30\,\mu$V to avoid heating effects. The output current from lead L$_2$ was detected by means of a current preamplifier in series with a lock-in amplifier. A blocking capacitor between the excitation source and L$_1$ was employed to remove any unwanted dc component of the bias. The conductance of one of the devices at 1.7 K as a function of the voltage applied to gates G$_2$ and G$_5$ (data not shown) displayed plateaus demonstrating ballistic transport across the ring which in turn made it possible to determine the number of one-dimensional subbands available for transport in each arm of the ring. Data presented in the following were collected in the regime where only one propagating subband is available in each arm. This choice maximizes the AB-oscillation contrast.

\begin{figure}[t]
\begin{center}
\includegraphics[width=8cm]{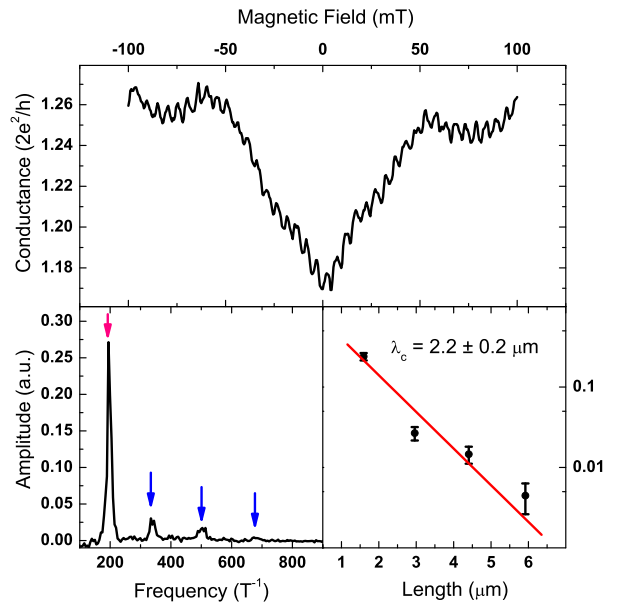}
\caption{(Color online) (a) Conductance AB oscillations measured at 400~mK setting the gates to: $V_1=V_3=V_4=V_6=0.5$~V and $V_2=V_5=0.8$~V; (b) Fourier transform (FT) of the AB oscillations. Average over the FT of different measurements is necessary to increase the peaks contrast, while derivative with respect to the magnetic field is applied to the AB oscillations to cut low frequencies contributions; (c) Exponential fit (solid line) of the FT peaks amplitudes (dots) as a function of the electron path length evaluated from the frequency of the FT peaks.}\label{fig2}
\end{center}
\end{figure}

The zero-bias conductance of device R500 measured at 400 mK as a function of the magnetic field is reported in Fig.~\ref{fig2}a. It displays oscillations with a period of $\sim 5.1$ mT as determined by Fourier transforming the data (see Fig.~\ref{fig2}b), and a contrast (defined as the peak-to-peak amplitude divided by the background value) of $\sim 1\,\%$. The observed period corresponds to $h/e$ AB oscillations for a ring with an effective radius of 507 nm, in good agreement with the sample geometry.
The higher-order harmonics appearing in the Fourier spectrum shown in Fig.~\ref{fig2}b can be exploited to estimate the electronic coherence length \cite{hansen_01}. The amplitude of each peak was evaluated with a Lorentzian-fit procedure. The center of each peak provides an estimate of the length of the corresponding closed electronic path. The exponential fit of the peak amplitude as a function of the electronic-path length yields a coherence length of $\lambda_c \sim 2.2\,\mu$m, a value consistent with what expected for a high-mobility 2DEG at 400 mK.

\begin{figure}[t]
\begin{center}
\includegraphics[width=8cm]{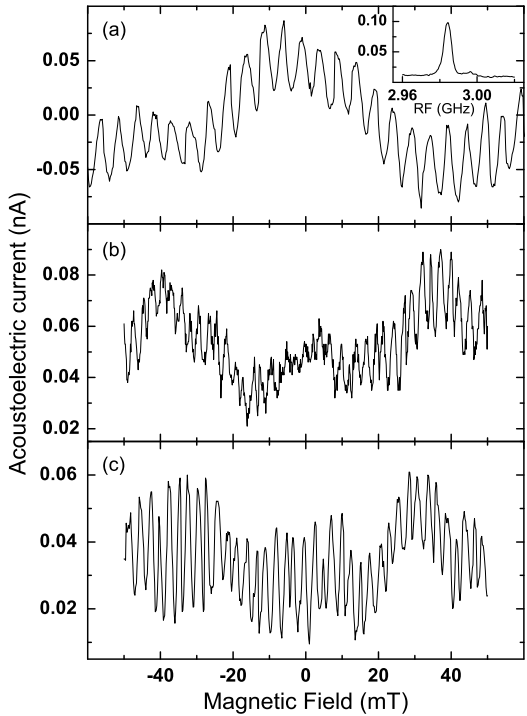}
\caption{Comparison between the Acoustoelectric currents measured at 400~mK in the two rings (\textbf{R500} and \textbf{R750}) as a function of the magnetic field.
(a) Acoustoelectric current in \textbf{R500}) generated by SAWs of 1~$\mu$m.
(b) and (c) are the acoustoelectric currents measured in \textbf{R750} and generated by SAWs of 1~$\mu$m and 2~$\mu$m wavelength.
In the inset we report the acoustoelectric current at $T = 1.7$ K, generated by a 1-$\mu$m-periodicity IDT across \textbf{R500} as a function of the RF excitation frequency. The excitation signal was modulated with pulses with a width of 300~ns and a period of 2500~ns.}\label{fig3}
\end{center}
\end{figure}

In order to investigate the acoustoelectric regime SAWs were generated by exciting the IDT incorporated in each device. Experiments were performed in the pulsed regime to avoid spurious effects due to cross-talk and SAW reflections off sample edges \cite{astley_06, astley_06b}. In the present case, the effects of cross talk and reflections were found to be minimized by using a pulse period of 2500 ns and pulse width of 300 ns. Acoustoelectric current across R500 measured at 1.7 K is shown as a function of the excitation frequency $f$ in the inset of Fig.~\ref{fig3}. The peak at $f = 2.985$ GHz corresponds to SAW generation at IDT resonance.

The acoustoelectric current was measured as a function of the magnetic field at T = 400 mK. Results obtained on R500 and R750 are presented in Figs.~\ref{fig3}a, \ref{fig3}b and \ref{fig3}c. Both devices exhibit pronounced oscillations. The periods of the observed AB oscillations correspond to ring radii of $505 \pm 10$ nm and $710 \pm 10$ nm for R500 and R750 respectively. These values match within experimental error the measured lithographic radii of the devices (500 nm and 715 nm respectively) and confirm that the origin of the observed oscillations is the magnetic AB effect. To the best of our knowledge, this is the first demonstration of the coherence of SAW-driven electronic transport in AB rings. 
By applying the same procedure adopted to analyze the linear-transport regime, data yield a coherence-length value of $\sim 2.4\,\mu$m, consistently with the linear regime.
Note that although the experimental acoustoelectric coherence length is comparable to the linear-transport coherence length, the contrast of the AB oscillations in the two different transport regimes is very different, being of the order of $1 \%$ in the latter case and $exceeding$ $100 \%$ in presence of SAWs.

This peculiar phenomenology can be clarified and successfully modeled by taking into account the nature of acoustoelectric transport in ballistic devices. Following Ref.~\cite{Totland96}, SAWs can be schematized as a monochromatic phonon flux characterized by momentum  $q=2\pi / \lambda_{SAW}$ and energy $q v_{SAW}$. Piezoelectric interactions between electrons confined in a one-dimensional channel and SAWs generate the acoustoelectric current due to momentum transfer to electrons that can be introduced in the system Hamiltonian as a first-order perturbation term. The solution of Boltzmann equation demonstrates that electron-phonon interaction is more efficient for electrons propagating in the same direction of the acoustic wave (set positive for clarity) and that a particularly strong electron-phonon coupling occurs when the SAW velocity matches the Fermi velocity in one of the subbands of the 1D channel ($v_{SAW} = v_{F}= \hbar k_{F} / m^* $). In the case of GaAs devices this condition occurs close to a subband pinch-off, leading to ``giant'' acoustoelectric current peaks \cite{Shilton96}. Experimental data \cite{Shilton96} also show that even off-resonance acoustoelectric current is detectable, even if lower in intensity owing to the weaker coupling.
This ``off-resonance'' regime is the one relevant for the present analysis since data were collected by biasing gates in order to operate the device at the first conductance plateau. In this situation the Fermi energy ($E_F$) can be estimated $E_F \sim 3$~meV based on the measured charge density, corresponding to a Fermi velocity ($v_F = \sqrt{2E_{F}/m^{*}} \sim 10^5$~m/s), i.~e.~two orders of magnitude higher than the SAW velocity ($v_{SAW} \simeq 3000$~m/s) in GaAs.
In this regime electron Fermi momentum $k_F \simeq \sqrt{2m^*E_F}/ \hbar \sim 10^8$ m$^{-1}$ is one order of magnitude higher than the phonon momentum $q= 2 \pi / \lambda_{SAW} \sim 10^7 $~m$^{-1}$, therefore non allowing phonon-induced backscattering of electrons. Moreover, electron-phonon interactions are, at first approximation, restricted to electrons 
within $E_F - \Delta E$ and $E_F + \Delta E$, where
$\Delta E = \max(\Delta E_{ph}, k_B T)$. Here $\Delta E_{ph}$ is the energy gained by an electron after absorbing a phonon, i.~e.~$\hbar q v_{SAW}$.
In our case $\Delta E_{ph} \simeq 10\,\mu$eV, corresponding to a temperature of $\simeq 0.1$ K. 

Under these conditions, Pauli exclusion principle makes the SAW-phonon absorption more efficient for the electrons propagating in the same direction of the SAW \cite{gurevich}. 
This leads to a larger perturbation of the Fermi distribution of the these electrons  ($f^{>}$), with respect to the distribution of electrons propagating in the opposite direction ($f^{<}$, here for simplicity we shall neglect this perturbation): $ f^{>}  =f_0 + \Delta f$, $f^{<}  \simeq f_0$,
where $f_0$ is the equilibrium Fermi function while $\Delta f$ represents the change in $f_0$ induced by the SAW. Here, $\Delta f$ is non-negligible only in the range $E_F + \Delta E > E  > E_F - \Delta E$ and assumes positive (negative) values at energies higher (lower) than $E_F$. Conservation of electron number implies that $\Delta f$ have zero average ($\int^{+\infty}_{0} \Delta f dE = 0$). The resulting acoustoelectric current generated across a constriction can then be written as:
\begin{equation}
\begin{split}
\label{AcEquation}
I_{ac} =
\frac{2e}{h} \int^{+\infty}_{0} (f^{>}-f^{<}) \mathbf{T}(E) dE \simeq \\
\frac{2e}{h}  \int^{E_F + \Delta E}_{E_F - \Delta E} \Delta f \mathbf{T}(E) dE,
\end{split}
\end{equation}
where energies are measured from the bottom of the first 1D subband and $\mathbf{T}(E)$ is the electron transmission probability across the constriction, accounting for interference effects.

The resulting integral, defined over an energy window $\Delta E$, suppresses the Fourier components of $\mathbf{T}(E)$ that oscillate with frequency much higher than $\Delta E$. Neglecting their contribution to the integral and keeping only low-periodicity components of $\mathbf{T}(E)$, the acoustoelectric current can be written:

\begin{equation}
\label{AcEquation2}
I_{ac} \simeq
\frac{2e}{h} \int^{E_F + \Delta E}_{E_F - \Delta E} \Delta f \mathbf{\overline{T}}(E) dE .
\end{equation}

where $\mathbf{\overline{T}}(E)$ includes the low-frequency Fourier components of the transmission probability.

This equation can be further simplified by substituting $\mathbf{\overline{T}}(E)$ with its first-order approximation around $E_F$ since, by definition, $\mathbf{\overline{T}}(E)$ varies slowly in the integration domain. 
\begin{equation}
\begin{split}
I_{ac}\simeq \frac{2e}{h} \int^{E_F + \Delta E}_{E_F-\Delta E} \Delta f \left[\mathbf{\overline{T}}(E_F)+ \frac{\partial \mathbf{\overline{T}}}{\partial E}(E_F)(E-E_F) \right] dE = \\
=\frac{2e}{h}  \frac{\partial \mathbf{\overline{T}}}{\partial E}(E_F) \int^{E_F + \Delta E}_{E_F- \Delta E} \mid (E-E_F) \Delta f \mid dE .
\end{split}
\end{equation}

The acoustoelectric current can therefore assume negative values depending on the slope of $\mathbf{\overline{T}}(E)$ consistently with the acoustoelectric counterflow shown in Fig.~\ref{fig3}.
It is instructive to compare this result to the expression for conductance given by the Landauer-B\"uttiker formalism at low temperatures:
$$
\sigma  \simeq \frac{2e^2}{h}  \mathbf{T}(E_{F}) \int^{+\infty}_{0} (-\frac{\partial f}{\partial E})dE
.$$
The reason for the high-visibility of acousoelectric AB oscillations is immediately apparent. In fact, owing to the derivative with respect to $E$, the slowly-varying, non-coherent contribution is not present in the acoustoelectric current so that $\frac{\partial \mathbf{\overline{T}}}{\partial E}(E_F)$ shows AB oscillations with a visibility higher than $\mathbf{T}(E_{F})$ even if both regimes are governed by the same coherence length.

In conclusion we should like to remark that the present model not only clarifies the peculiar behavior of acoustoelectric coherent transport, but establishes an unexpected link between acoustoelectric and thermoelectric transport regimes.
This relationship is clear if we compare acoustoelectric current given by (Eq.~\ref{AcEquation}) and the thermoelectric coefficient $B$:

$$B(T) = \frac{2e}{h} \frac{1}{T} \int^{+\infty}_{0} (-\frac{\partial f}{\partial E}) (E-E_F) \mathbf{T}(E) dE .$$

Since the $(-\frac{\partial f}{\partial E}) (E-E_F)$ is functionally analogous to the perturbation $\Delta f$ defined in an energy range $\Delta E = k_B T$ around $E_F$, in the weak coupling regime described by Eq.~\ref{AcEquation2}, the acoustoelectric current is proportional to the thermoelectric coefficient at temperature $T= \Delta E /k_B$:

$$ I_{ac} \propto  B (T= \Delta E /k_B) .$$

This finding suggests that the study of thermoelectric effects in nanoscopic circuits can be successfully carried out with the much simpler experimental approach characteristic of acoustoelectric investigations.

\end{document}